\documentclass[sigconf,nonacm,natbib=false]{acmart}

\AtBeginDocument{%
  \providecommand\BibTeX{{%
    \normalfont B\kern-0.5em{\scshape i\kern-0.25em b}\kern-0.8em\TeX}}}

\copyrightyear{2020} 
\acmYear{2020} 
\setcopyright{acmcopyright}
\acmConference[KDD '20]{Proceedings of the 26th ACM SIGKDD Conference on Knowledge Discovery and Data Mining}{August 23--27, 2020}{Virtual Event, CA, USA}
\acmBooktitle{Proceedings of the 26th ACM SIGKDD Conference on Knowledge Discovery and Data Mining (KDD '20), August 23--27, 2020, Virtual Event, CA, USA}
\acmPrice{15.00}
\acmDOI{10.1145/3394486.3403287}
\acmISBN{978-1-4503-7998-4/20/08}

\graphicspath{ {./img/} }
\usepackage{subfig}
\usepackage{balance}
\acmSubmissionID{861}

\usepackage[style=ACM-Reference-Format,backend=bibtex,sorting=none]{biblatex}
\begin{document}
%\fancyhead{}

\title{Cellular Network Radio Propagation Modeling with Deep Convolutional Neural Networks}

\author{Xin Zhang}
\authornote{Both authors contributed equally to this work.}
\authornote{The authors completed this work in 2018 when they were with Futurewei Technologies, Inc.}
%%\email{xinzhang75@gmail.com}
\affiliation{%
  \institution{Futurewei Technologies, Inc.}
  \city{Bridgewater}
  \state{New Jersey}
  \country{USA}
}
\author{Xiujun Shu}
\authornotemark[1]
\affiliation{%
  \department{GTS R\&D Algorithm Dept.}
  \institution{Huawei Technologies, Co., Ltd.}
  \city{Shenzhen}
  \country{China}
}
\author{Bingwen Zhang}
\authornotemark[2]
\affiliation{%
  \institution{Futurewei Technologies, Inc.}
  \city{Bridgewater}
  \state{New Jersey}
  \country{USA}
}
\author{Jie Ren}
\affiliation{%
  \department{GTS R\&D Algorithm Dept.}
  \institution{Huawei Technologies, Co., Ltd.}
  \city{Shenzhen}
  \country{China}
}
\author{Lizhou Zhou}
\affiliation{%
  \department{GTS R\&D Algorithm Dept.}
  \institution{Huawei Technologies, Co., Ltd.}
  \city{Shenzhen}
  \country{China}
}
\author{Xin Chen}
\affiliation{%
  \department{GTS R\&D Algorithm Dept.}
  \institution{Huawei Technologies, Co., Ltd.}
  \city{Shenzhen}
  \country{China}
}

\renewcommand{\shortauthors}{Xin Zhang, et al.}

\begin{abstract}
Radio propagation modeling and prediction is fundamental for modern cellular network planning and optimization. Conventional radio propagation models fall into two categories. Empirical models, based on coarse statistics, are simple and computationally efficient, but are inaccurate due to oversimplification. Deterministic models, such as ray tracing based on physical laws of wave propagation, are more accurate and site specific. But they have higher computational complexity and are inflexible to utilize site information other than traditional global information system (GIS) maps. 

In this article we present a novel method to model radio propagation using deep convolutional neural networks and report significantly improved performance compared to conventional models. We also lay down the framework for data-driven modeling of radio propagation and enable future research to utilize rich and unconventional information of the site, e.g. satellite photos, to provide more accurate and flexible models.  \end{abstract}


%\ccsdesc[500]{General and reference~General conference proceedings}


\maketitle
\pagestyle{empty}

\section{Introduction}
\label{intro}
In a wireless communication system, signal transmitted by a base station can undergo attenuation as it propagates through space before reaching a mobile user equipment (UE). The difference in strength between the transmitted and received signal is commonly referred to as path loss, usually measured in dB. 

Path loss may be due to many effects, such as, free-space loss, refraction, diffraction, reflection, aperture-medium coupling loss, and absorption. Path loss can be influenced by terrain contours, environment (e.g., urban or rural, vegetation or foliage, and the like), propagation medium (e.g., dry or moist air), the distance between transmitter and receiver, the height and location of antennas, the antenna radiation pattern, and other factors. 

Path loss is often used in analysis and design of link budget of a telecommunication system, and is fundamental for modern cellular network planning, a process to optimize the number of base stations, the locations of these base stations, antenna selection and parameter settings, etc, such that maximum coverage is achieved with minimal cost. 

Measuring path loss at all locations of interest is expensive, and sometimes impossible when planning an area without existing base stations. Therefore, it is desirable to predict path loss at a mobile UE using radio propagation models without actual measurement. Conventional radio propagation models are categorized into two types: empirical and deterministic models \cite{antenna_prop4_wireless_book,sarkar03_survey}. 

Empirical models describe the relationship between path loss and environment parameters using multiple equations, which are determined empirically using coarse statistics of data measured under a limited number of settings, specified by environment parameters. Empirical models are computationally efficient, but are usually inaccurate because they use only very limited parameters to describe the environment.

Deterministic models on the other hand describe the environment using detailed information such as GIS maps, including building locations and shapes, terrain of the area, land use (clutter layer), and antenna settings (location, azimuth, tilt and radiation pattern). These models are derived from physical laws of wave propagation, and in theory could be computed exactly by solving Maxwell’s equations. Unfortunately, this approach requires unrealistically complex mathematical operations and computing power. Ray tracing technique is a popular approximation using geometrical optics and knife edge diffraction theory \cite{ikegami91_ray,walfisch88_ray}. Deterministic models are usually more accurate than the empirical ones, at the cost of more computation. Furthermore, actual path loss is determined by much richer information about the environment than conventional GIS maps can provide, so the prediction by deterministic models still has considerable error.

Radio propagation modeling is essentially a regression problem in machine learning terms, and quite a few works have proposed to use machine learning methods, mainly artificial neural networks, to predict path loss \cite{mgbe15_ann,sotiroudis13_ann,sotiroudis15_ann,popoola18_ann,mom2014_ann,ayadi17_ann}. These works propose to use regression neural network to predict path loss based on limited description of the environment and are usually not site specific, similar to empirical models. Fully connected neural networks with one or a few hidden layers are proposed, and the input information are usually numerical vectors with 10 to 20 dimensions, only able to capture very coarse characteristics of the environment. 

Recently deep convolutional neural networks (ConvNet) have significantly outperforms other machine learning techniques in computer vision related tasks, such as image recognition \cite{lenet_cnn,alexnet_cnn,vgg_cnn,inception_net_cnn,resnet_cnn} and image segmentation \cite{fullyconv15_cnn,hariharan_cnn,Ronneberger15_cnn}. These deep architectures have exponentially more representation power and significantly outperform conventional shallow learning methods. Usually much more training data is also necessary for them to work properly. 

In this article, a data-driven modeling framework is proposed to reformulate path loss prediction to an image regression problem, and utilize recent breakthrough of deep convolutional neural networks in computer vision related tasks. We demonstrate how input information to the model, such as maps and antenna parameters, can be translated into image tensors. We also proposed a specific design of deep convolutional neural network architecture that is tailored for radio propagation modeling. Results from large amount of simulated and real field data show that the proposed data-driven model performs significantly better than conventional radio propagation models. 

It is also worth pointing out that the framework we proposed is flexible and can accommodate a variety of future designs. One aspect is novel input information design. For example, building materials and vegetation have large effect in high frequency radio propagation, but they are expensive if not impossible to include in conventional GIS maps. On the other hand, both building materials and vegetation can be observed in satellite photos and can be easily included into input tensors. Another aspect of future design is researchers can invent better neural network architectures for more accurate models. 

 The contribution of this work lies in the following aspects:
 \begin{itemize}
   \item{We reformulate the radio propagation modeling to an image regression problem and propose to use deep ConvNet to solve it.}
   \item{We apply radio propagation domain knowledge and design input tensors to capture most relevant input information.}
   \item{We select suitable neural network architectures based on the nature of radio propagation modeling, as demonstrated in Section \ref{network_arch}.}
   \item{We tested the proposed model using both simulated and real field data to show significant superiority over conventional models, and the proposed model has already been deployed as a Huawei internal tool and is being used by field engineers over the world.}
 \end{itemize} 

\section{Conventional Radio Propagation Models}
\subsection{Empirical Models}

There are a number of commonly used empirical models, for example, Okumura-Hata model \cite{hata80_emp}, Stanford University Interim model \cite{stanford_interim_emp}, standard propagation model (SPM) \cite{spm_emp} and Walfisch-Ikegami model \cite{antenna_prop4_wireless_book,sarkar03_survey}. 

The above models all use a few parameters and equations to characterize typical classes of radio links such as urban, suburb and rural. The parameters are determined based on measured and averaged losses in the aforementioned typical classes of ratio links. Thus empirical models are not site-specific, i.e., they use exactly the same parameters for two different sites as long as the sites fall in the same link category. 

For example, the Okumura-Hata model is valid for microwave frequencies from 150 to 1500 MHz, and is suited for both point-to-point and broadcast communications, and covers mobile station antenna heights of 1-10 m, base station antenna heights of 30-200 m, and link distances from 1-10 km. The Okumura-Hata model uses separate sub-models for the urban, suburban and rural environments. The urban model is formulated as follows. 
\begin{equation}
\begin{split}
L_U = & \; 69.55 + 26.16\log_{10}f - 13.82\log_{10} h_B \\
&-C_H + [44.9 - 6.55\log_{10}h_B] \log_{10}d
\label{HataLU}
\end{split}
\end{equation}
For small or medium-sized city, 
\begin{equation}
C_H = 0.8 + (1.1\log_{10}f - 0.7)h_M - 1.56\log_{10}f
\end{equation}
For large cities and carrier frequency in $150\text{MHz} \le f \le 200\text{MHz}$,
\begin{equation}
C_H = 8.29(\log_{10}(1.54h_M))^2 -1.1
\end{equation}
And for large cities and carrier frequency in $200\text{MHz} \le f \le 1500\text{MHz}$,
\begin{equation}
C_H = 3.2(\log_{10}(11.75h_M))^2 -4.97
\end{equation}
$L_U$ is the path loss in urban areas in decibel (dB). $h_B$ is the height of base station antenna in meter. $h_M$ is the height of mobile station antenna in meter. $C_H$ is the antenna height correction factor.
$f$ is the carrier frequency in Megahertz.
$d$ is the distance between the base and mobile stations in kilometer. 

The models for suburban and rural areas are just slightly modified version of the above model and are omitted here.

SPM, derived from Okumura-Hata model with a few more parameters, is another commonly used model in industry, and we will compare our proposed method with SPM in Section \ref{sec_experiments}.

As we can see, the disadvantage of empirical models is that they use very limited input information and parameters to coarsely characterize radio links, so the prediction can be very inaccurate. In our experience, the root-mean-square error (RMSE) between the un-calibrated prediction and ground truth is typical around $12$-$15$dB.

\subsection{Deterministic Models}
\begin{figure}[htbp]
	\centerline{\includegraphics[width=\columnwidth]{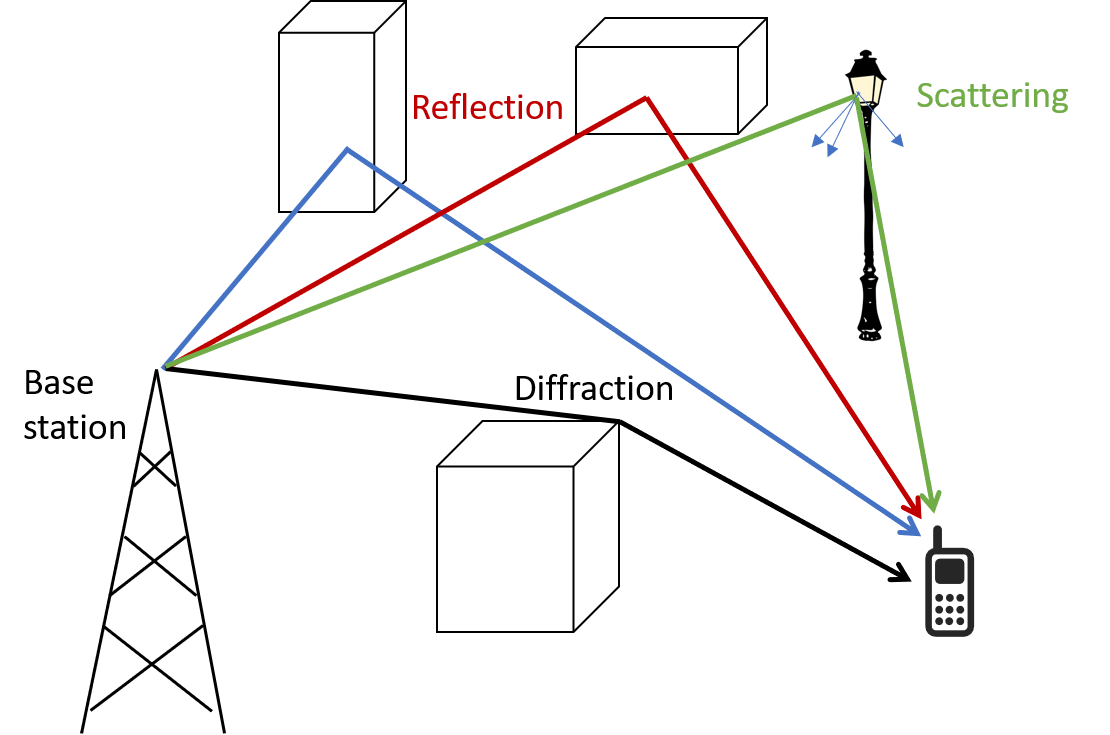}}
	\caption{Ray tracing model.}
	\label{fig_ray_tracing}
\end{figure}

Ray tracing techniques based on geometrical optics \cite{ikegami91_ray,walfisch88_ray} produce the most widely used deterministic path loss models. Reflection, scattering and diffraction are modeled and calculated using ray tracing, and the effects are combined as shown in Figure \ref{fig_ray_tracing}. 

The input to ray tracing models contains much more information about a specific site than that of empirical models. GIS maps, often containing building layers (specifying heights, shapes and locations of buildings), terrain layers (altitude of ground) and clutter layers (categorized land use), are typically used as input to ray tracing. Examples of these layers can be found in Figure \ref{fig_tensor_layers}, where the values in each layer are typically normalized.

Antenna parameters such as location, height, azimuth, tilt and radiation pattern are also required in ray tracing.

Ray tracing can calculate deterministic prediction of path loss at any specified points. With site-specific and rich input information, it is usually more accurate than empirical models. However, many factors affecting radio propagation such as the material of building enclosure and density of vegetation are not available in typical GIS maps, resulting in prediction error (measured in RMSE) close to 10dB. Thus, to perform well for a specific site in practice, ray tracing models require calibration which is essentially over-fitting a few parameters to measurements from the same site using expensive drive tests. Moreover, in a new site planning task, these measurements are outright not available. 

Ray tracing based radio propagation models are very complex and are implemented as commercial softwares by several companies implementing their proprietary technologies. They are usually very computationally intensive and are often too slow for cellular network planning, which is essentially an optimization process with thousands of steps and each step recalculates the path loss of the whole area under consideration. 

Another disadvantage of ray tracing models is that unconventional input information such as satellite photos of areas cannot be easily incorporated into the models. These new sources contain even richer information than traditional GIS map layers.  

\section{Proposed Neural Network Architecture and Training}

In this section we present the details of proposed data driven propagation model, including input tensor design, neural network architecture and training methods. 

\subsection{Problem Formulation}
\label{sec_problem_formulation}

We observe that the inputs to ray-tracing based propagation models include GIS map layers, which are essentially numerical matrices and can be considered as the channels of an image, as shown in Figure \ref{fig_tensor_layers}. Each pixel of the image here is corresponding to a square area patch of $m$ by $m$ meters in the physical world, where $m$ is the resolution of GIS map. Typical resolutions of an GIS map are 5 meters, 10 meters and 20 meters.

Ray-tracing based models also utilize engineering parameters such as antenna location (latitude, longitude and height), orientation and radiation pattern as input information. These parameters are not in matrix format and cannot be directly incorporated into input image tensors. There are two possible ways to include these inputs into machine learning frameworks, e.g., neural network architectures. We can keep the original format of these engineering parameters and use as scalar inputs to a neural network in parallel to the image tensor constructed with GIS map layers. By this design we are asking the neural network to learn some underlying mechanisms on these parameters which we already know with simple physics. We think a better approach is to translate these parameters into additional channels of the input image tensor so that we have a unified input in image format. By these translations, we are essentially applying domain knowledge into the input tensor design, as will be demonstrated in the next subsection. 

Propagation models predict path loss values at each location in an area, which can be gridded into square patches of $l$ by $l$ meters. If we assume the path loss at any point within a patch to be the same (a common practice by field engineers), the output can be formatted into a matrix of path loss values, which can also be considered as an image with each pixel corresponding to a small area patch of $l$ by $l$ meters. The intensity of a pixel is set to the path loss value at the square patch representing the pixel. Note that in practice $l$ is typically larger than or equal to $m$. 

Thus the prediction process is essentially an image-to-image regression problem, i.e., mapping from one image domain to another. Instead of using ray tracing techniques to calculate the mapping, we propose to use an alternative machine learning and data driven approach to learn the mapping. More specifically we proposed to use a deep convolutional neural network (ConvNet) as the propagation model and use simulated and real field data to train the ConvNet. Details of the network architecture and training methodology will be presented later in this section.  

\subsection{Input Tensor Design}
\label{sec_tensor_design}

One key aspect to use a ConvNet as propagation model is the design of input tensor, such that all necessary information is organized in a format to be utilized effectively by the neural network.
\begin{figure*}[ht]
	\centering
	\includegraphics[width=\textwidth]{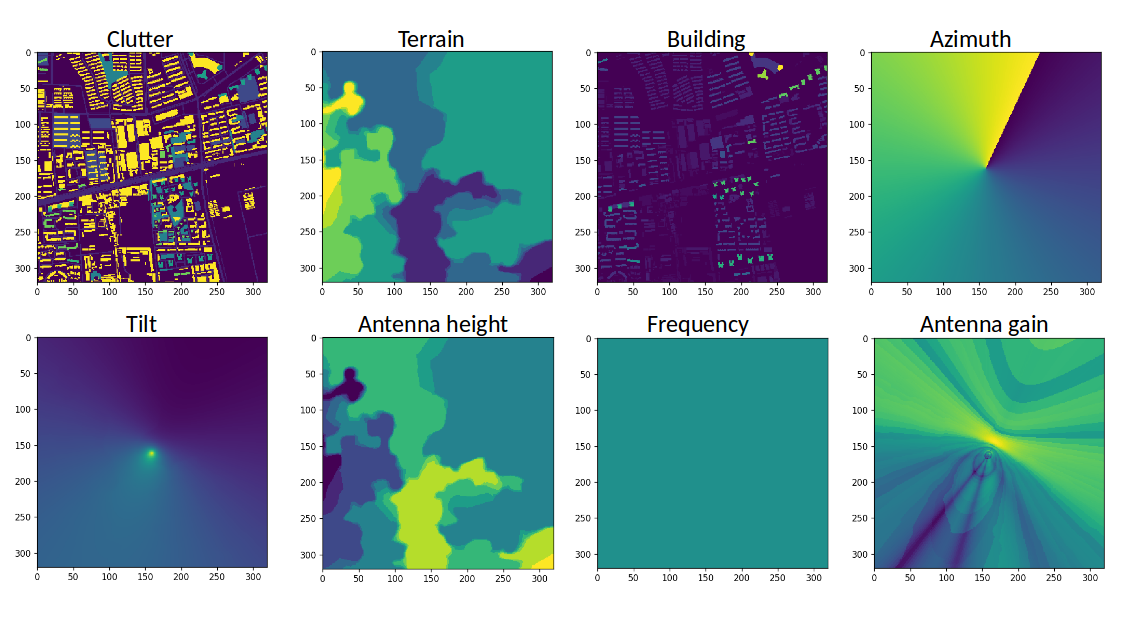}
	\caption{Channels of input tensor to PLNet.}
	\label{fig_tensor_layers}
\end{figure*}

We design the input to be a 3-dimensional tensor, with the shape of $C\times W \times H$. $C$ is the number of channels defined below. $W$ and $H$ are the width and height (in pixels) of the target area and they are design parameters. For example, if the GIS map has a resolution of $m=5$ meters and we would like to predict the path loss within an area of 1000 by 1000 meters, the width and height of the patch are both 200 pixels. 

As mentioned previously, layers (building, terrain and clutter) of GIS maps are naturally image channels, while engineering parameters have to be translated into extra channels of the input image tensor. 

Instead of using extra channels to encode the latitude and longitude of the antenna, we propose to use the 2-D location of the antenna as the center to cut out a map patch (with all three layers) from the whole map. Subsequent channels also use the same center.  

In practice the antenna can be either installed on a small tower on top of a building, or hang on the side. We encode the antenna height information in a channel of the input tensor where each pixel has the same value representing the antenna height in meters. 

With the above intuition, we design the input tensor with 8 channels and summarize each channel as follows:
\begin{itemize}
	\item \textbf{Clutter} channel contains the clutter layer of a GIS map, which categorizes each pixel into one of the surface feature types, such as sea, forest, urban, river, lake, village and other ground cover information. It is a categorical value and usually numericalized as integers from 1 to 21. 
	\item \textbf{Building} channel contains the building layer of a GIS map, which describes the general outline and height information of buildings. If there is building on a pixel, the value of the pixel is the height of building at the pixel. If there is no building on the pixel, the value is simply zero.
	\item \textbf{Terrain} channel contains the terrain layer of a GIS map, describing the altitude of each pixel, which is normalized by subtracting the lowest point in the map patch. 
	\item \textbf{Azimuth} channel indicates the horizontal angle difference between the line of sight from each pixel to the antenna and direction of main lobe.
	\item \textbf{Tilt} channel indicates the vertical angle difference between the line of sight from each pixel to the antenna and direction of main lobe.
	\item \textbf{Antenna height} channel is designed to encode the height of antenna simply with the height in meters at each pixel of the map patch.  
	\item \textbf{Frequency} channel describes the frequency of carrier signal. Signals with different carrier frequencies usually have different attenuation, hence different path losses. We simply copy the frequency value of a sample (the carrier frequency of the base station) to each pixel. 
	\item \textbf{Antenna Gain} channel describes the antenna radiation pattern, i.e., how the radio energy is distributed in the space around antenna. We assume the 3-D radiation pattern is intercepted by the terrain surface below the antenna, and the gain at each pixel on the surface is filled in this channel of the tensor. 
\end{itemize}
Examples of the channels are shown in Figure \ref{fig_tensor_layers}. Note that in this framework, the input tensor can be easily modified when new information is available. For example, satellite photos of an area can be incorporated as an additional channel for the input tensor.

\begin{figure*}[ht]
	\centerline{\includegraphics[width=\textwidth]{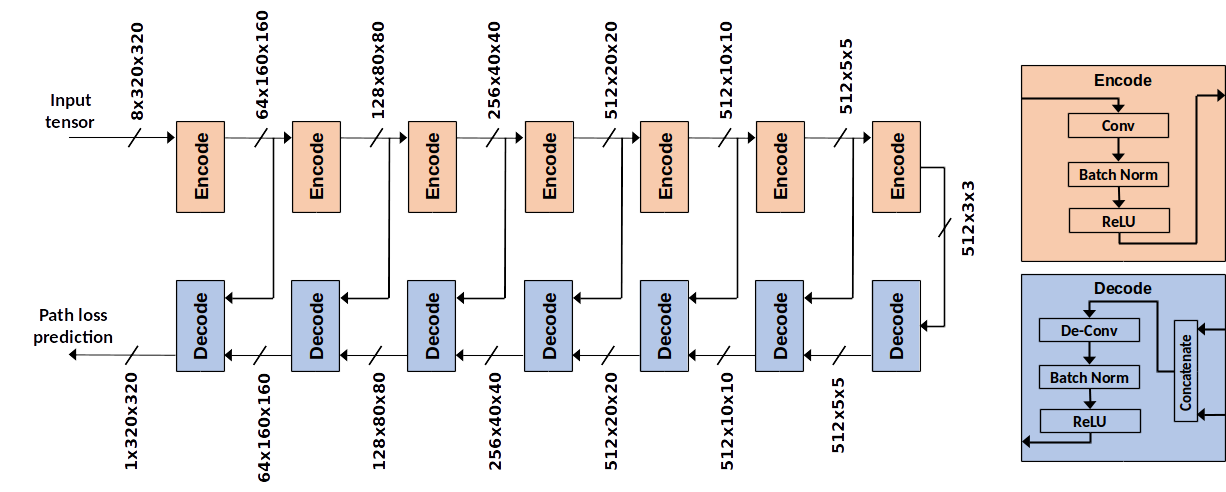}}
	\caption{Architecture of PLNet.}
	\label{fig_architecture}
\end{figure*}

\subsection{Network Architecture}
\label{network_arch}
Referring to the conventional propagation models, we observe that the path loss at a specific location is determined by both the macro condition of the environment such as the overall density of buildings in the area, and the micro condition close to the receiver such as the number of walls nearby and their specific locations. 

To capture the above characteristics, a good neural network architecture for path loss prediction should assemble different levels of abstraction inside a neural network to form predictions. We propose to use a U-Net \cite{Ronneberger15_cnn} type of architecture, which is also known as encoder-decoder with skip connections. These skip connections are critical and serve as the combinations of various abstraction levels. The detailed architecture is shown in Figure \ref{fig_architecture}.

In all the convolution and de-convolution layers, we use 3x3 kernel size and set stride to 2. Padding is applied such that feature map size is only altered by striding. Note that in Figure \ref{fig_architecture} the input tensor was designed to have the same width and height as the output tensor (path loss matrix), i.e., the input map layers has the same resolution as the path loss prediction. If they have different resolution, it is easy to modify the architecture to up-sample/down-sample the path loss matrix in the width and height dimensions. We dub our neural network PLNet for easy reference later in this article. 

\subsection{Supervised Training}
\label{sec_supervised_training}
The most straight forward way to train PLNet is supervised training. The input tensors are based on engineering parameters of base stations and GIS maps, constructed according to Section \ref{sec_tensor_design}. Corresponding labeled path loss matrices are collected from field measurements or alternatively from commercial path loss simulators. Details about data generation will be explained in Section \ref{sec_experiments}. With the labeled path loss matrices (supervisors) for all the training samples, we can use mean absolute error (MAE) or mean square error (MSE) as the loss function for training. MAE is defined as follows and MSE can be defined similarly,
\begin{equation}
\begin{split}
\mathcal{J}_{L1}\left(\mathbf{\Theta}\right) 
& =\mathbb{E}_{\mathbf{(X,Y)}}\left\|\mathbf{Y} - 
\hat{\mathbf{Y}}\left( \mathbf{X}, \mathbf{\Theta}\right) \right\|_{1} \\
& \approx \frac{1}{NWH}\sum_{i=1}^{N} \sum_{j=1}^{W}\sum_{k=1}^{H} \left|\mathbf{Y}_i(j,k)-\hat{\mathbf{Y}}_i(j,k)\right|
\end{split}
\label{L1-cost-function}
\end{equation}
where $\mathbf{\Theta}$ is the weights of PLNet and $\mathbf{X}=\{\mathbf{X}_i\}_{i=1}^N$ are the input tensors as described in Section \ref{sec_tensor_design}. $\mathbf{Y}=\{\mathbf{Y}_i\}_{i=1}^N$ and $\hat{\mathbf{Y}}=\{\hat{\mathbf{Y}}_i\}_{i=1}^N$ are the ground-truth and predicted path loss matrices respectively. $N$ is the number of samples, and $W$ and $H$ are the width and height of the path loss matrix. $\mathbf{Y}_i(j,k)$ refers to path loss value at the $j^{th}$ row and $k^{th}$ column of the ground truth path loss matrix of the $i^{th}$ sample.

Note that in field collected data not all pixels have valid measurements, because it is not realistic to ask road testing equipment to traverse every single pixel in the area. Actually it is not uncommon that only 5-10\% of the pixels in a path loss matrix have measured values. Pixels without valid measurements should be excluded from the calculation of loss function, hence will not affect the gradient back propagation. Just clamping the error from these pixels at zero will effectively prevent them from interfering the training process.

\section{Experiments}
\label{sec_experiments}
\begin{figure*}[ht]
	\centering
	\subfloat[Ground truth (simulated by ray tracing)]{\includegraphics[width=0.77\columnwidth]{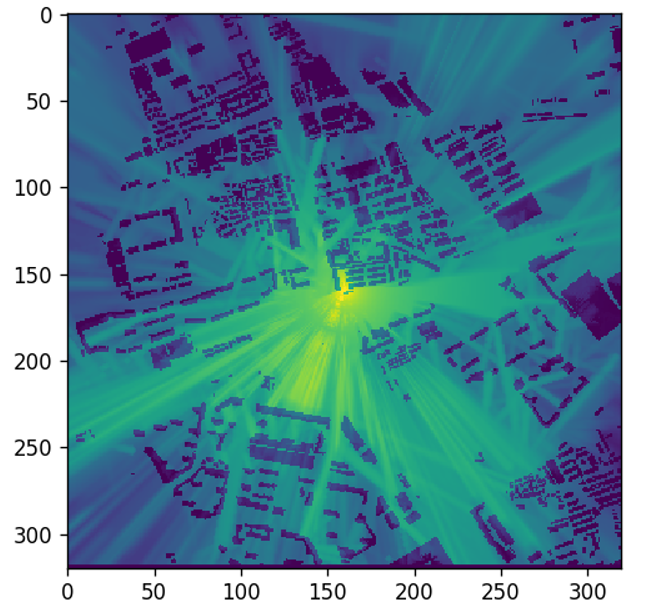}}
	\hfil
	\subfloat[Predicted by PLNet]{\includegraphics[width=0.9\columnwidth]{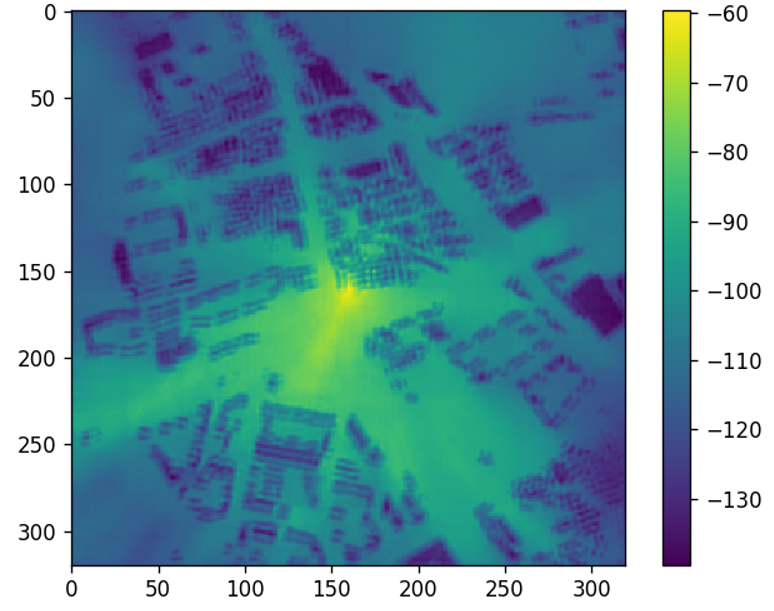}}
	\caption{Path loss matrices of ground truth calculated by commercial simulator compared with PLNet prediction.  }
	\label{fig_simulated_coverage}
\end{figure*}

The proposed method is tested with both simulated data and real field data. We use simulated data in addition to real field data for two reasons. 

First, it takes time and effort to collect significant amount of field data to train PLNet from scratch, and using simulated data to pre-train PLNet can benefit prediction accuracy when only small amount of real data is available. In our case, the pre-training is effective because, in both the simulated and real data, the input tensors have the same distribution, i.e., they are both constructed using GIS map layers, a set of know antenna radiation patterns, and same ranges of antenna parameters. The only difference between the simulated and real data is the path loss values (supervisors), so simulated data are especially useful to train lower level representations in PLNet.

The second reason is that field data are usually proprietary and there is no open data set for path loss prediction problem. For other researchers to test our proposed method, GIS map layers and commercial path loss simulators (ray tracing based) are available to the public (some may charge license fees). The simulator we used in generating simulated data is the ``Volcano propagation model" developed by Siradel SAS \cite{volcano_ray}. 

\begin{figure*}[ht]
	\centering
	\subfloat[Building channel]{\includegraphics[width=0.9\columnwidth]{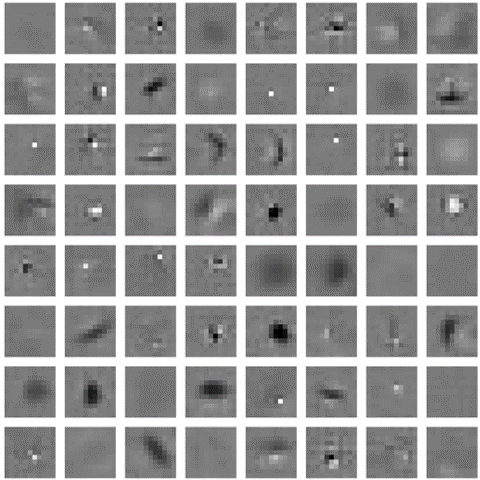}}
	\hfil
	\subfloat[Terrian channel]{\includegraphics[width=0.9\columnwidth]{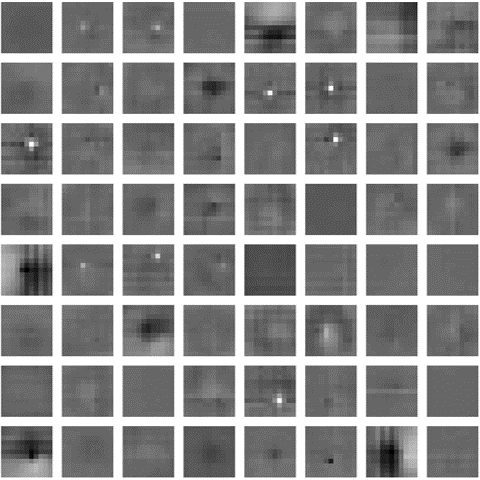}}
	\caption{Filters of the first convolution layer of PLNet. The weights of filters corresponding to the building and terrain channels are plotted. }
	\label{fig_first_layer_filters}
\end{figure*}
To generate simulated data, We use GIS maps of 40 cities around the world, and randomly select suitable locations to ``install" (virtual) antennas. Note that the selected antenna locations need to be far enough from each other to ensure independence among samples. We also select antenna radiation patterns randomly from a set of known antennas, and set the antenna parameters such as azimuth, tilt and transmitting power randomly. Given the GIS maps and antenna parameters, Volcano simulator is used to calculate the corresponding path loss matrices. We managed to generate more than 800,000 simulated samples.

We also collected about 400,000 field data samples using drive test or mobile measurement report (contains GPS locations and received signal power) from more than 50 cities worldwide. As pointed out in Section \ref{sec_supervised_training}, not all pixels have valid measurements, so we adjust the cost function and back propagation accordingly as described in Section \ref{sec_supervised_training}.

We increased the effective number of both simulated and field samples by data augmentation techniques such as reflection and rotation.

Root-mean-square error (RMSE) is used as the metric for prediction accuracy and it is calculated by averaging over all pixels and samples.

\subsection{Results with Simulated Data}
First we clarify that the results from this subsection, i.e., results using simulated data (generated from ray tracing based models), are to demonstrate that PLNet has the capacity to capture meaningful patterns from the input tensor by "learning from ray tracing". In real application, PLNet will compete with ray tracing based models, and we show in the next subsection that PLNet trained by real field data outperformed ray tracing significantly. 

We use the simulated data from 36 cities (728,574 samples) to train PLNet, and data from the other 4 cities (96,724 samples) to test the performance of PLNet. 

It takes about 80 epochs for the training to converge, and we achieve test performance of $\text{RMSE}=7.48\text{dB}$ using data from the 4 test cities. 

Samples of predicted and ground truth (calculated by Volcano in this case) path loss matrices are shown in Figure \ref{fig_simulated_coverage}. The predicted path loss matrix captures the overall patterns of the ground truth, but it is a little blurred as a result of using MSE/MAE as loss function.

We also visualize the convolutional filters of the first layer in Figure \ref{fig_first_layer_filters}. The original PLNet has $3\times3$ convolution filters for each channel and it is hard to visualize such small filters, so we increase the filter size of the first layer filters to $11\times11$ just for visualization purpose. We can see that the filters for building and terrain channels are indeed trained to discover low level patterns of the corresponding map layers. 

There are two interesting observations. First, the filter patterns of the terrain channel is wider and smoother than those of the building channel. This is because buildings have much sharper edge in map compared to smooth changing terrains. Second, the filter patterns are similar to the ConvNets trained for image recognition purpose, with the exception of some patterns having distinctive highlighted pixels. When these filters convolve with the input building or terrain channels, they search for higher buildings or terrains near the current location. This is because protruding objects can reflect electromagnetic waves and affect radio propagation.

From these results, we can see that the lower layers of PLNet are trained to capture patterns of input map, even though it is trained with simulated data generated by raytracing models. Thus it can be a good pre-training in the case where real field data is limited. 

\subsection{Results with Field Data}
Field collected data from 48 cities world-wide (319,350 samples) are utilized to train our PLNet, and data collected from another 4 cities (8,316 samples) are used to test the performance of PLNet against empirical and ray-tracing based methods. The results are shown in Table \ref{tbl_field_result_table}, where PLNet ST stands for PLNet by supervised training, and SPM is a widely applied empirical model in industry. We can see the proposed PLNet out-perform the conventional methods by a large margin in each test city. 

\begin{table}[htbp]
	\caption{RMSE (in dB) of proposed methods compared with conventional methods using real data. Note that we anonymized the city names for privacy concerns.} % title of Table
	\centering % used for centering table
	\begin{tabular}{c c c c c} % centered columns (5 columns)
		\hline\hline %inserts double horizontal lines
		City & PLNet ST & Ray Tracing & SPM \\ [0.5ex] % inserts table
		%heading
		\hline % inserts single horizontal line
		City A & 8.44 & 11.84 & 16.56 \\ % inserting body of the table
		City B & 9.54 & 12.87 & 17.15 \\
		City C & 9.04 & 12.28 & 16.88 \\
		City D & 8.81 & 12.02 & 17.02 \\ [0.5ex] % [1ex] adds vertical space
		\hline %inserts single line
	\end{tabular}
	\label{tbl_field_result_table} % is used to refer this table in the text
\end{table}

\begin{figure*}[htbp]
	\centering
	\begin{tabular}{ccc}
		\subfloat[Field ground truth]{\includegraphics[width=0.8\columnwidth]{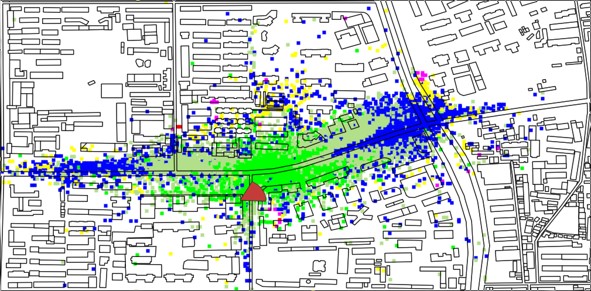}} 
		& \subfloat[Prediction by PLNet]{\includegraphics[width=0.8\columnwidth]{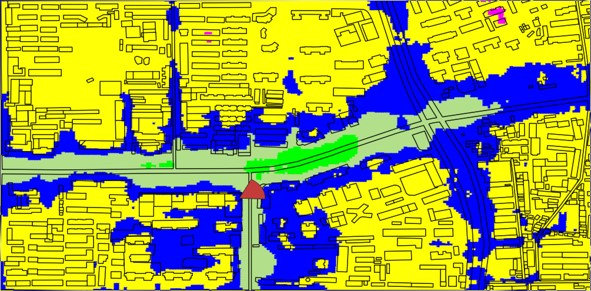}} & \\
		\subfloat[Prediction by commercial ray tracing simulator]{\includegraphics[width=0.8\columnwidth]{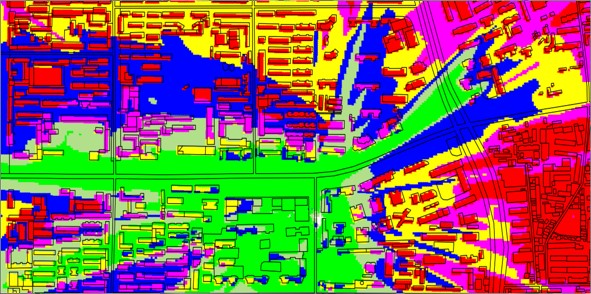}} 
		& \subfloat[Prediction by SPM]{\includegraphics[width=0.8\columnwidth]{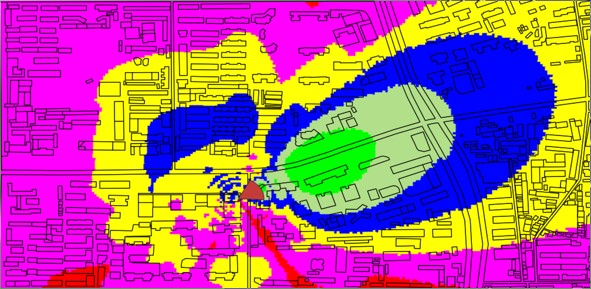}} & \subfloat[Legend]{\includegraphics[width=0.25\columnwidth]{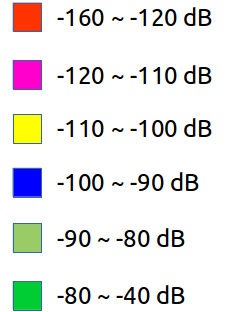}}
	\end{tabular}
	\caption{Predicted path loss matrices by PLNet and conventional methods. }
	\label{fig_real_data_PLmatrix_samples}
\end{figure*}

We also plot samples of predicted path loss matrices by PLNet and conventional methods in Figure \ref{fig_real_data_PLmatrix_samples}. Note that only the center area of the ground truth path loss matrix has valid values. PLNet predictions are more consistent with ground truth and are less dramatic in the surrounding area.

Because of the large errors, both empirical and ray-tracing based methods recommend to ``calibrate" their models before actual prediction. Basically calibration is overfitting the conventional models to certain region of consideration, for example a city. It is worth pointing out that in many cases data required by calibration are not available, for example, in new construction of cellular networks, where no base station has been constructed yet. In other cases, data required by calibration has to be recorded through expensive road test. So PLNet has considerable advantage here because it can perform well without calibration. 

\begin{figure}[htbp]
	\centerline{\includegraphics[width=0.9\columnwidth]{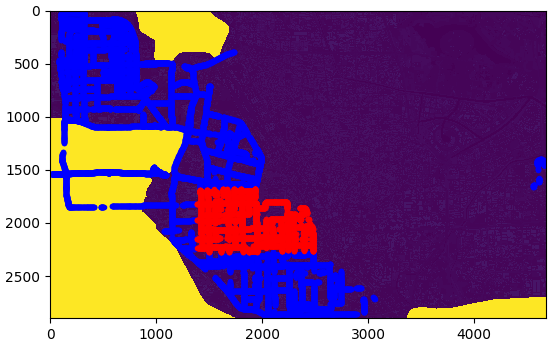}}
	\caption{Map of City D where the calibration data is collected from the roads in blue color, and the data collected from roads in red color are used for test after the calibration.}
	\label{fig_zhengzhou_map}
\end{figure}

But for the sake of completeness, we present the performance comparison after calibration. A map of City D is shown in Figure \ref{fig_zhengzhou_map}, where the calibration data is collected from the roads in blue color, and the data collected from roads in red color are used for test after the calibration. 

Both SPM and Volcano use calibration data to fine-tune a few parameters designed for calibration \cite{volcano_ray}. For PLNet, we use the calibration data to fine tune the neural network for a few epochs. The RMSE for SPM and Volcano are 9.50dB and 8.53dB respectively, improved from the un-calibrated performance. PLNet still has the best accuracy with $\text{RMSE}=7.81\text{dB}$. 

PLNet also has significant computation speed advantage over ray tracing based method. To predict the path loss matrix of a single cell with $320\times320$ pixels, Volcano needs 30 seconds (8-core Intel i7 CPU), while PLNet only needs 0.33 seconds (implemented in TensorFlow with one Nvidia Titan Xp GPU). Of course PLNet benefits from GPU computation, but at this time we are not aware of any ray tracing based path loss prediction software taking advantage of GPU acceleration.  

\section{Conclusions}
In this paper we presented a new method to predict the radio propagation 
path loss using a deep convolutional neural network, which we dub as PLNet. We designed the input tensor to PLNet to include detailed GIS map layers and antenna parameters. We also proposed network architecture of PLNet that are tailored for this application. Using real field data, we showed that the proposed method is significantly more accurate than conventional methods and can also be computed much faster using modern GPUs. 

\begin{acks}
The authors would like to thank Dr. Jin Yang and Dr. Yan Xin for inspiring discussions during the course of this work. The authors would also like to thank the anonymous referees for their valuable comments and helpful suggestions. 
\end{acks}

%\bibliographystyle{ACM-Reference-Format}
%\bibliography{paper-kdd}

\balance
\printbibliography


\end{document}